# Low-Thermal-Budget Ferroelectric Field-Effect Transistors Based on CuInP$_2$S$_6$ and InZnO

Hojoon Ryu,[1] Junzhe Kang,[1] Minseong Park,[2] Byungjoon Bae,[2] Zijing Zhao,[1] Shaloo Rakheja,[1] Kyusang Lee[2,3] and Wenjuan Zhu*[1]

[1]Electrical and Computer Engineering, University of Illinois at Urbana-Champaign, Urbana, IL 61801, USA,

[2]Electrical and Computer Engineering, University of Virginia, VA 22904, USA

[3]Material Science and Engineering, University of Virginia, VA 22904, USA

ABSTRACT: In this paper, we demonstrate low-thermal-budget ferroelectric field-effect transistors (FeFETs) based on two-dimensional ferroelectric CuInP$_2$S$_6$ (CIPS) and oxide semiconductor InZnO (IZO). The CIPS/IZO FeFETs exhibit non-volatile memory windows of ~1 V, low off-state drain currents, and high carrier mobilities. The ferroelectric CIPS layer serves a dual purpose by providing electrostatic doping in IZO and acting as a passivation layer for the IZO channel. We also investigate the CIPS/IZO FeFETs as artificial synaptic devices for neural networks. The CIPS/IZO synapse demonstrates a sizeable dynamic ratio (125) and maintains stable multi-level states. Neural networks based on CIPS/IZO FeFETs achieve an accuracy rate of over 80% in recognizing MNIST handwritten digits. These ferroelectric transistors can be vertically stacked on silicon CMOS with a low thermal budget, offering broad applications in CMOS+X technologies and energy-efficient 3D neural networks.

**Introduction**

Emerging non-von Neumann architectures with intensive in-memory computing will demand high-density integration of embedded memory [1, 2]. Three-dimensional monolithic integration of logic transistors and memory among interconnects will enable us to expand the on-chip memory density [3-5]. These architectures will not only overcome the limitation of chip area in two-dimensional (2D) layout but also facilitate the development of new three-dimensional (3D) computation systems. In these systems, logic and memory elements are closely intertwined, resulting in significant improvements in memory access bandwidth and energy efficiency. Among various non-volatile memories, ferroelectric memory stands out as one of the most promising candidates due to its low-power consumption, high endurance and long retention [6-9].

However, to fully realize a 3D system comprising ferroelectric memories and silicon CMOS, several fundamental technology challenges need to be addressed. Traditional ferroelectric memories primarily rely on complex perovskites, such as lead zirconate titanate (PZT) and strontium bismuth tantalate (SBT) [10, 11]. These ferroelectric materials are typically grown through epitaxy on a single-crystal substrate, which presents a significant challenge for integrating them with silicon CMOS due to the lattice match restriction. Furthermore, PZT and SBT exhibit short retention times and face limitations in terms of thickness scaling [12]. In recent years, doped hafnium oxide has emerged as a new class of ferroelectric materials [13-19]. Ferroelectric hafnium oxide offers several advantages, including a high coercive field, excellent scalability down to 2.5 nm, and good compatibility with CMOS processing [19-22]. However, the realization of the ferroelectric phase transition in doped hafnium oxide requires high-temperature annealing. For Al-doped $HfO_2$, annealing temperatures ranging between 650 and 1000 °C are necessary [16, 23], while for Zr-doped $HfO_2$, the typical annealing temperature is around 500 ºC [24]. High-temperature annealing can introduce significant reliability issues for the underlying silicon CMOS and Cu/low-k interconnects. In this study, we address this issue by employing van der Waals (vdW) ferroelectrics, specifically $CuInP_2S_6$ (CIPS), in the memory devices. CIPS can be stacked on any material and does not necessitate high-temperature annealing. Moreover, 2D CIPS exhibits a low coercive field and lacks dangling bonds at the interface, making it an ideal candidate for high-speed memory transistors that operate at low voltages [25-27].

Another challenge in 3D integration involves synthesizing low-thermal-budget semiconducting materials with high carrier mobility and good uniformity. Various types of semiconducting materials have been studied to assess their compatibility with back-end-of-line (BEOL) processes. Amorphous silicon and organic semiconductors can be processed at low temperatures but exhibit low carrier mobility [28-30]. Polycrystalline silicon can offer higher carrier mobility but is more expensive to fabricate and prone to non-uniformity [31]. In recent years, 2D materials have been extensively explored due to their atomic thin body, good carrier mobility and ability to be transferred onto any substrate at room temperature [32, 33]. However, the wafer-scale synthesis and defect-free transfer of 2D materials remain to be fully developed. Oxide semiconductors, which can be deposited at low temperatures with wafer-scale coverage and exhibit high carrier mobility, hold great promise for BEOL memory devices [34-38]. Among various oxide semiconductors, indium-zinc-oxide (IZO) is particularly attractive as a channel material due to its high carrier mobility, low off-state current, low-temperature processability, and good uniformity [39-41]. In this work, ferroelectric field-effect transistors (FeFETs) based on CIPS/IZO heterostructures are demonstrated with a processing temperature of 250 °C. These CIPS/IZO FeFETs exhibit high carrier mobility (37 $cm^2$/V-s) and low off-current (3.3x$10^{-8}$ µA/µm), as well as a large memory window (~ 1 V). Furthermore, the neural networks based on CIPS/IZO FeFETs demonstrate more than 80% accuracy in recognizing MNIST handwritten digits. These ferroelectric device will have high broad applications in neuromorphic computing and CMOS+X technology.

**Results and discussion**

The ferroelectricity of CIPS was measured on metal/CIPS/metal capacitors. Figure 1(a) shows the optical image of a CIPS capacitor, with the active area of CIPS capacitor marked. The thickness of the CIPS is 250 nm, as shown in Figure 1(b). In addition, the CIPS flakes were characterized by Raman spectroscopy, as depicted in Figure 1(c). The presence of distinct and identifiable Raman peaks in the CIPS material provides evidence supporting the existence of the ferroelectric phase of CIPS. We observed the Raman peak at 266 cm$^{-1}$ can be attributed to the vibrations involving sulfur (S), phosphorus (P), and sulfur (S) atoms (S-P-S vibrations) [26, 42]. On the other hand, the peak observed at 376 cm$^{-1}$ can be attributed to the stretching of phosphorus-phosphorus (P-P) bonds ) [26, 42]. In Figure 1(d), a symmetric hysteresis loop is observed at room temperature with a remanent polarization of 4.4 μC/cm$^2$, which is consistent with the previous findings on CIPS [25, 43, 44]. To measure the ferroelectric hysteresis loop, triangular read pulses were employed, as illustrated in the inset of Figure 1(d). The preset pulse has the same rise/fall time as the read pulse but possesses an opposite polarity as the first read pulse. The remnant polarization was measured using a Keithley semiconductor parameter analyzer (model 4200-SCS) equipped with a 4225-PMU ultra-fast I-V module.

Following the characterization of the CIPS capacitor, we investigated the FeFET based on CIPS/IZO structure. The CIPS/IZO FeFETs were fabricated using the following process. First, Ni back gates with 20 nm thickness were formed on the 280 nm SiO$_2$/Si substrates. Then we deposited a 20 nm thick $HfO_2$, serving as the back gate dielectric, using atomic layer deposition (ALD). Then, we used optical lithography to pattern the channel area. The 10 nm IZO channel was deposited on the $HfO_2$/Ni stack using a sputter tool, followed by annealing at 250 °C in the air to remove the chemical residue. The source/drain electrodes (10 nm Ti/30 nm Au) were then formed by optical lithography, metal evaporation, and lift-off. Next, we transferred the ferroelectric CIPS on the IZO channel. This CIPS layer plays two roles in the device: (1) serves as the ferroelectric layer to tune the carrier density of the IZO channel, and (2) acts as the passivation layer of the IZO channel to block the $O_2$ and moisture in the air. Ni/Au (30/90 nm) top gate was deposited on CIPS in the last step. Here, the channel length is 16 µm, and the channel width is 43 µm. The schematic of the device structure and the optical image of the FeFET are shown in Figure 2(a) and 2(b) respectively. The fabrication flow of the FeFET is shown in Figure 2(c).

Figures 2(d) show the transfer characteristics of the CIPS/IZO transistor. For the CIPS/IZO FeFETs, the measurements were conducted at room temperature in a vacuum chamber. The back gate voltage ($V_{BG}$) was applied to the Ni gate. The transfer curve of the CIPS/IZO FeFET exhibits negligible hysteresis, when $V_{BG}$ is swept from -7 V to 3 V. It indicates the absence of significant trap charges between $HfO_2$ and IZO. At a drain voltage ($V_D$) of 0.1 V, the CIPS/IZO FeFET exhibits on-state drain current ($I_{ON}$) of 0.51 µA/µm and off-state drain current ($I_{OFF}$) of $3.3 \times 10^{-8}$ µA/µm. The on-state drain current was extracted at a $V_{BG}$ of 1 V, while the average off-state drain current was calculated as the average drain current within the gate voltage range of -7 V to -6 V. Notably, the off-state current of the IZO-based transistor is significantly lower (10-100 times) than that of ferroelectric transistors based on 2D transition metal dichalcogenide (TMDC), silicon, or other metal-oxide semiconductors, as indicated in Table 1 [26, 45-49]. The gate leakage at a $V_{BG}$ of 3 V is less than $10^{-6}$ µA/µm. Our FeFET shows a sizable $I_{ON}/I_{OFF}$ current ratio of $1.5 \times 10^7$ under $V_D$ = 0.1 V. The field-effect mobility is extracted using the equation $\mu = (dI_{DS}/dV_{GS}) \times (L/(WC_iV_{DS}))$, where the $L$ is the channel length, $W$ is the channel width, and the $C_i$ is the oxide capacitance. The CIPS/IZO/HfO$_2$ stack shows a mobility of 37 cm$^2$/V-s, which is higher than the previous cases [48, 50-53], indicating that the CIPS passivation layer can preserve the carrier mobility of the IZO channel. Figure 2(e) shows the transistor's output characteristics. The linear output curves at low drain voltages indicate that the contacts are Ohmic-like.

The memory operation of these FeFETs was tested by applying program pulses between the Ni back gate and the top gate on CIPS. Figure 3(a) shows the transfer curves of a CIPS/IZO FeFET measured after the program and erase operations, showing a large memory window (~ 1 V). When the applied program pulse ($V_p$) exceeds the coercive field, the dipoles in CIPS can switch direction and align with the external electric field. These ferroelectric polarizations in CIPS can modulate the type and density of the carriers in the IZO channel, as illustrated in Figure 3(b). When a negative program pulse is applied on the back gate, the downward polarization in CIPS will attract electrons in the IZO layer, which leads to a negative shift of the threshold voltage. On the other hand, when a positive program pulse is applied on the back gate, the upward polarization in CIPS will attract holes in IZO and induce a positive shift of the threshold voltage, as shown in Figure 3(b). After the program or erase operation, the transistor was measured through the Ni/HfO$_2$ gate with a fixed gate voltage range. The reading voltage is kept small, and the top gate of CIPS is left floating to minimize polarization switching in CIPS during the read operation. The retention of the FeFET with programmed and erased states was measured by monitoring the drain current as a function of time, as shown in Figure 3(c). The ratio between the high $I_{DS}$ at the programmed state (on-state) and the low $I_{DS}$ at the erased state (off-state) are kept at ~10$^3$ over 1000 seconds, which indicates the stable memory operation of our FeFET.

Furthermore, we investigated the potential of using CIPS/IZO FeFETs as artificial synaptic devices for neural network applications. A series of program pulses with amplitude modulation was applied to the FeFET and the transfer curves were measured after each pulse. In Figure 4(a), with an increase in positive (negative) pulse amplitude, the threshold voltage of the FeFET shifts gradually to the positive (negative) direction, which can be attributed to the existence of multiple domains in CIPS. The partial polarization switching in CIPS leads to the intermediate states of the FeFET. Figure 4(b) illustrates the pulse scheme used in these measurements. The retention of these states is shown in Figure 4(c). After a program pulse was applied, the drain current at a constant $V_{BG}$ (-1.26 V) was measured as a function of time, as shown in Figure 4(c). More than ten distinguishable multistate levels are observed with good retention. Based on the measured drain current, we obtained the dependence of the channel conductance on the pulse numbers, as depicted in Figure 4(d). The dynamic range of the weight update, i.e., the ratio between maximum conductance ($G_{max}$) and minimum conductance ($G_{min}$), is 125. This value shows a higher dynamic range compared to other existing records, as presented in Table 1.

Finally, we simulated the neural networks based on CIPS/IZO FeFETs, as shown in Figure 5(a). A single-layer perceptron structure was used for the CIPS/IZO neural networks for fast classification and compact layout. We employ rectifying linear-unit (ReLU) as an activation function. The synaptic connections between the input and hidden layer are evolved with respect to the training epochs, and up to 80% recognition accuracy is achieved in the MNIST handwritten digit classification, as shown in Figure 5(b).

**Conclusions**

In this work, we demonstrate high-performance FeFETs based on CIPS and IZO at a process temperature as low as 250 ºC. The CIPS/IZO FeFET shows a ~1 V memory window, over $10^6$ $I_{ON}/I_{OFF}$ ratio, low off-state current, and high carrier mobility. We also demonstrate the CIPS/IZO FeFET as an artificial synaptic device for neural network applications. The CIPS/IZO FeFET shows a sizeable dynamic ratio (>100) and gradual weight update. The neural network based on CIPS/IZO FeFETs exhibits over 80% accuracy in recognition of MNIST handwritten digits. These FeFETs based on 2D CIPS/thin film IZO have a high potential for 3D heterogeneous integration and cognitive computing.

**Conflicts of interest**

There are no conflicts to declare.

**Acknowledgments**

The authors would like to acknowledge the support from the Semiconductor Research Corporation (SRC) under Grant SRC 2021-LM-3042.

AUTHOR INFORMATION

**Corresponding Author**

*wjzhu@illinois.edu

ACKNOWLEDGMENT

The authors would like to acknowledge the support from the Semiconductor Research Corporation (SRC) under Grant SRC 2021-LM-3042.

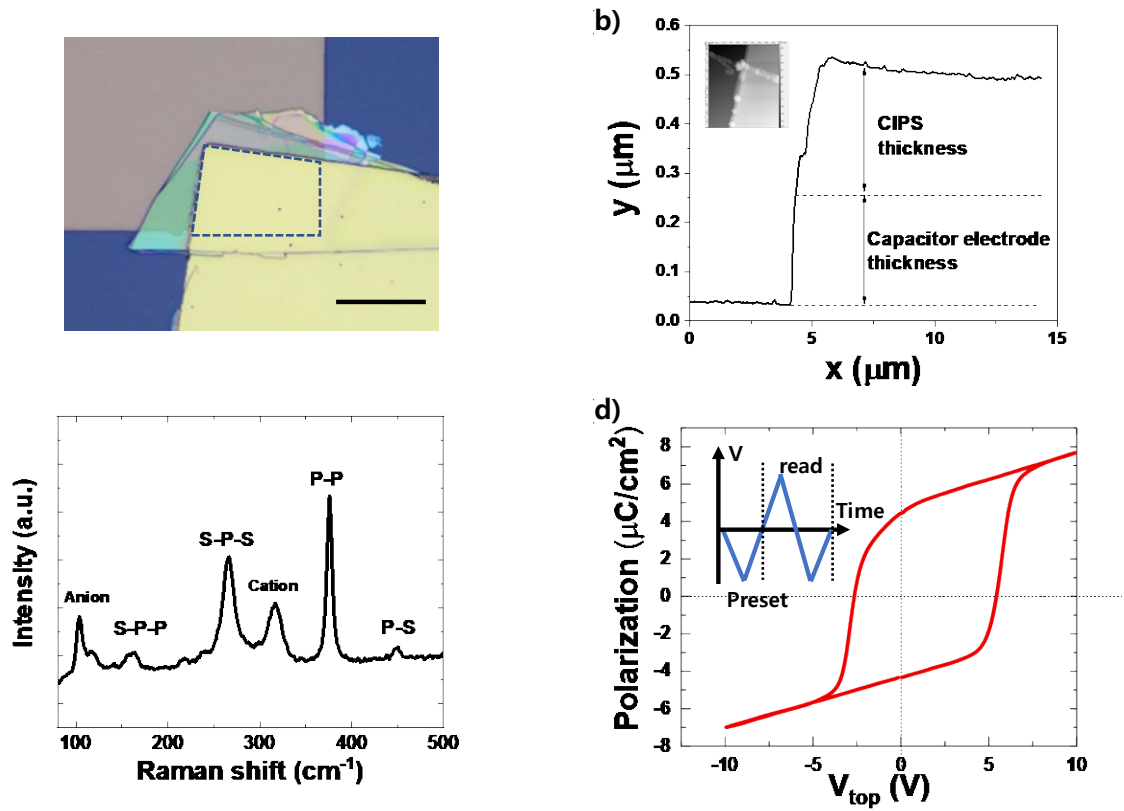

Figure 1. (a) Optical image of a metal/CIPS/metal capacitor. (b) Height profile of the 2D CIPS flake measured by AFM. The extracted thickness is ~250 nm and the average roughness is 0.8 nm. The inset shows the AFM topography image of the CIPS flake. (c) Raman spectra of a 2D CIPS flake. (d) Polarization-voltage (P-V) loop of ferroelectric CIPS capacitor. The inset shows the waveform of applied voltage pulse during the measurement.

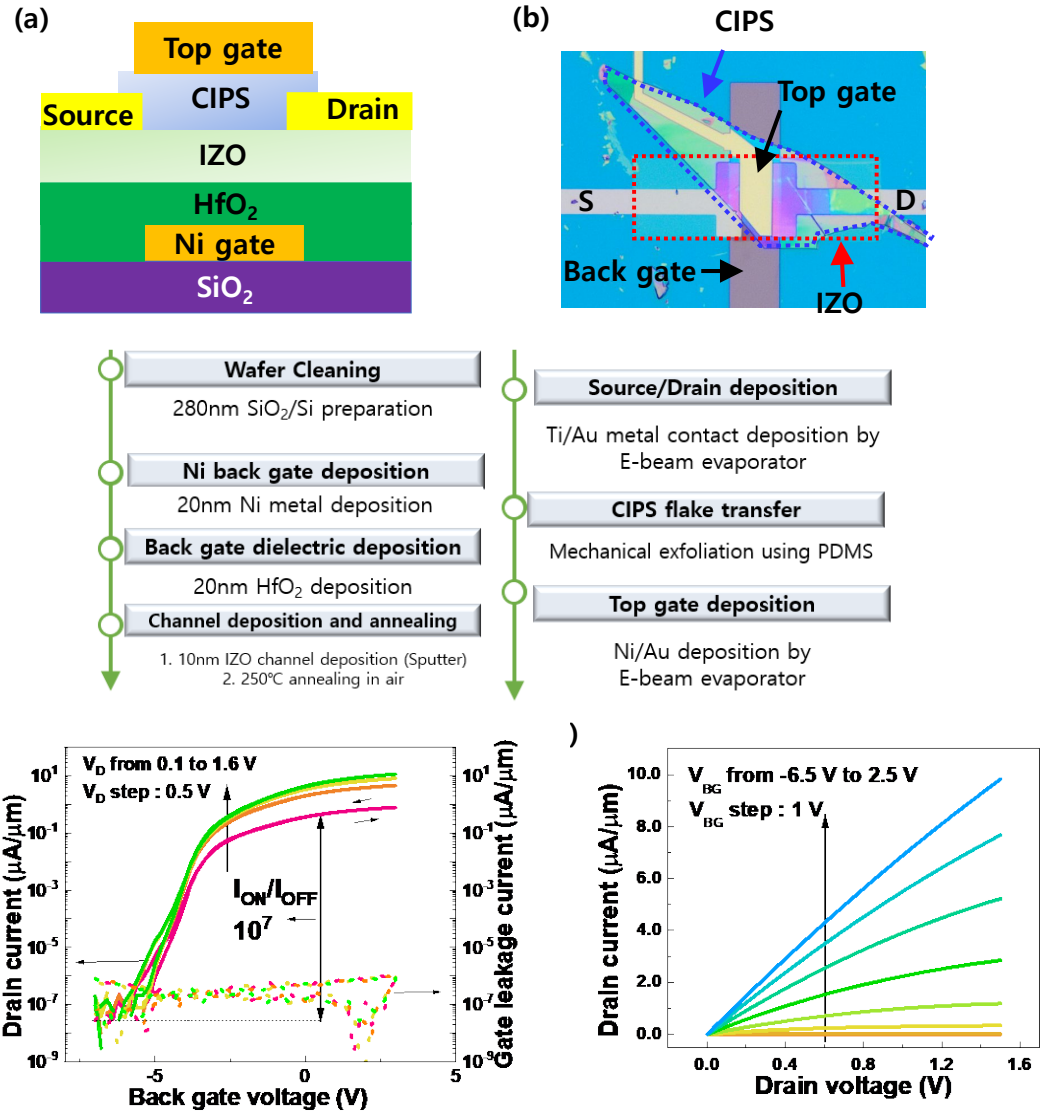

Figure 2. (a) Schematic and (b) optical image of a CIPS/IZO FeFET. (c) Process flow of the CIPS/IZO FeFET. (d) Transfer characteristics and (e) output characteristics of the CIPS/IZO FeFET.

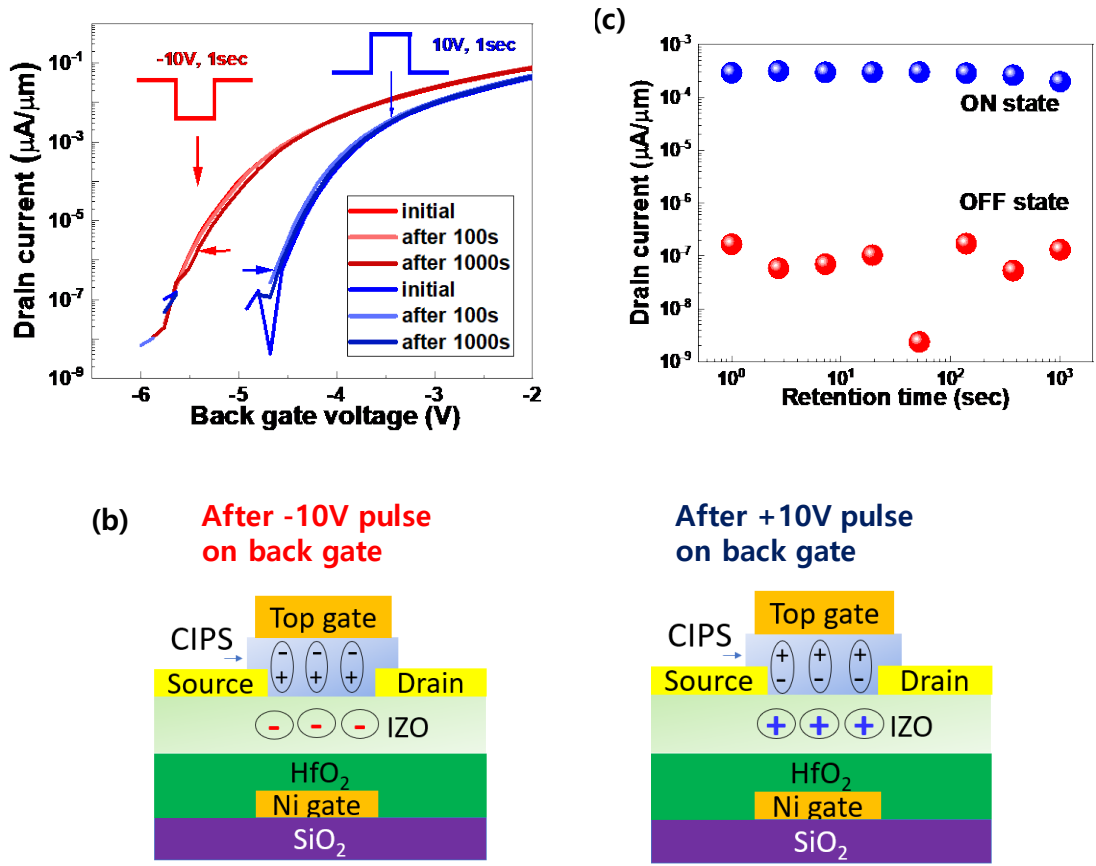

Figure 3. (a) The transfer curves of the CIPS/IZO based FeFET measured after applying program and erase pulses. (b) Illustration of the influence of ferroelectric polarization charges in CIPS on the carrier types in IZO channel. (c) The retention measurement of CIPS/IZO FeFET.

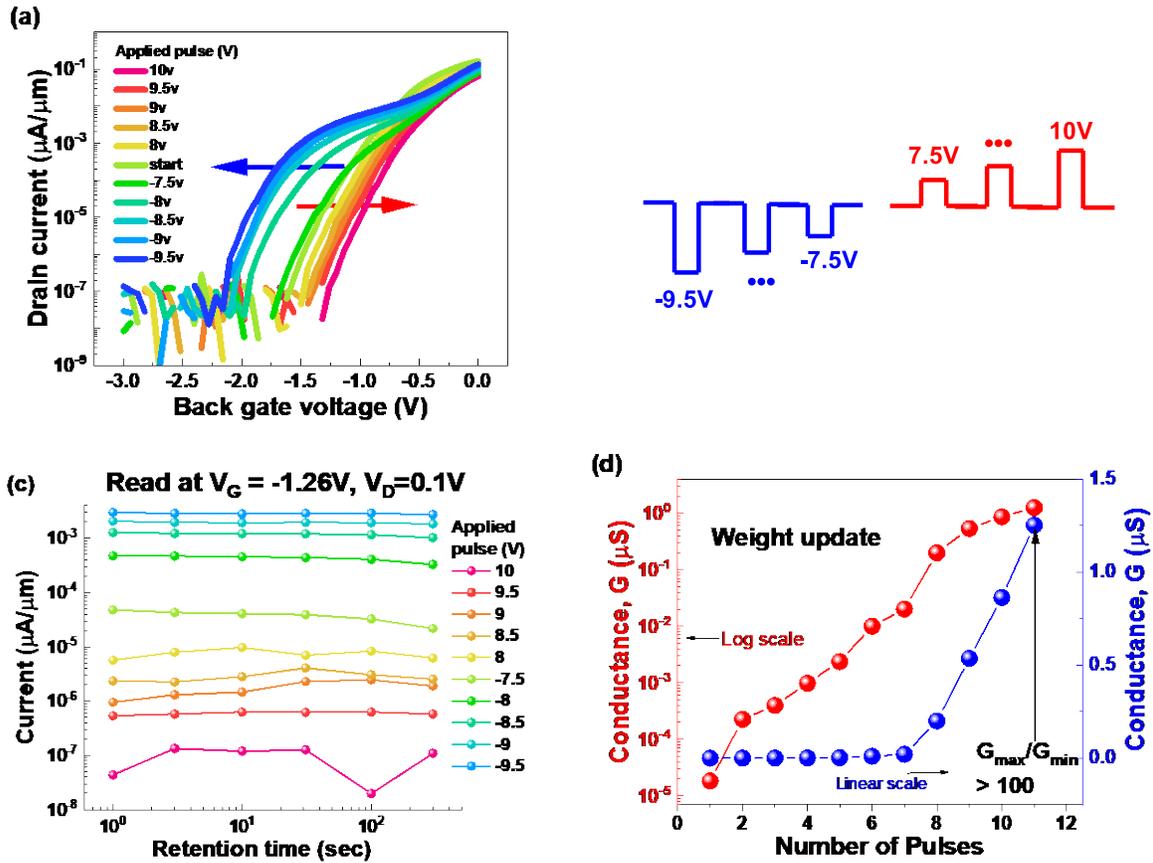

Figure 4. (a) The transfer characteristic of the CIPS/IZO FeFET exhibits a gradual shift following the application of sequential program pulses with pulse amplitude modulation. (b) The pulse scheme is depicted, demonstrating the modulation of pulse amplitudes. (c) Retention of the multi-states in the CIPS/IZO FeFET measured with various program amplitudes. (d) Conductance of the CIPS/IZO FeFET as a function of pulse number.

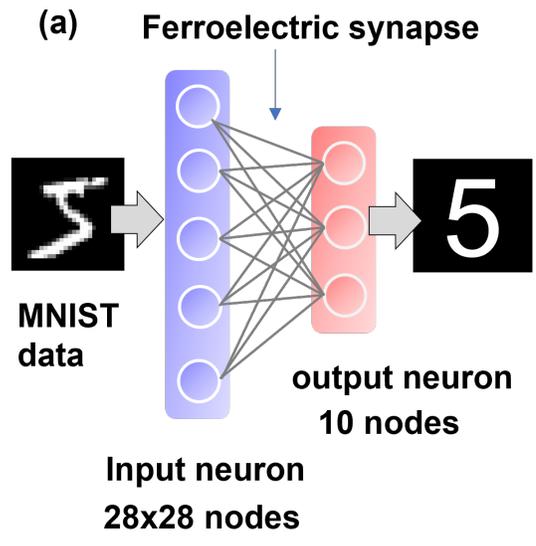

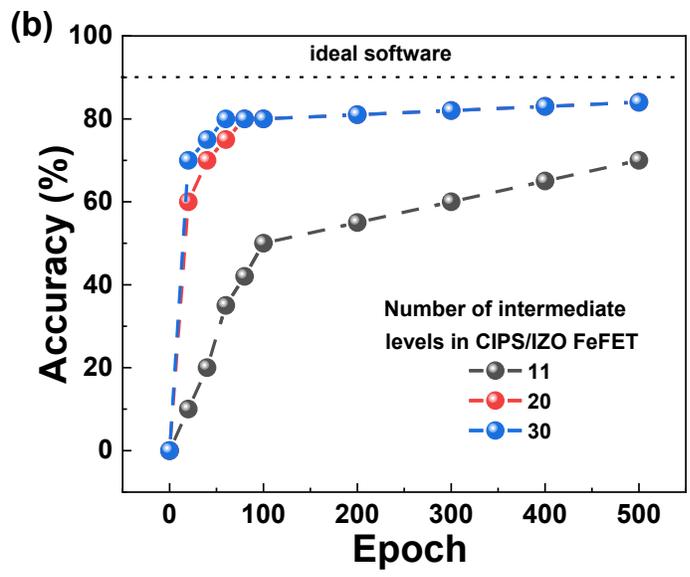

Figure 5. (a) Schematic of neural network and MNIST data set. (b) Recognition accuracy of the neural network based on the CIPS/IZO FeFETs.

TABLE I

SMALL CAPS: BENCHMARK TABLE OF FeFETs FOR MEMORY AND NEURAL NETWORK APPLICATIONS

| Device | P/E ratio | Memory window (V) | Retention (sec) | Weight update, $G_{max}/G_{min}$ ratio | Thermal budget (°C) |
|--------|-----------|-------------------|-----------------|----------------------------------------|---------------------|
| Our work | $>10^4$ | 1 (pulse) | $>10^3$ | >100 | 250 (BEOL compatible) |
| CIPS/WSe$_2$ [29] | $>10^5$ | 4.2 ($V_G$ sweep) | 50 | - | Room temperature |
| CIPS/MoS$_2$ [17] | $10^4$ | 2 ($V_G$ sweep) | - | - | Room temperature |
| HZO/IZTO [29] | $>10^3$ | 1.5 ($V_G$ sweep) | - | 6.5 | 400-500 |
| AlScN/MoS$_2$ [30] | $10^4$ | 20 ($V_G$ sweep) | $>10^4$ | - | 400 |
| PZT/IGZO [31,32] | $10^3$ | 2 ($V_G$ sweep) | $>10^3$ | >100 | >600 |

* The P/E ratio is the ratio of the measured drain current at program and erase states.